\documentclass[%
 reprint,
superscriptaddress,
 amsmath,amssymb,
 aps,
prb,
floatfix,
]{revtex4-1}

\usepackage{graphicx}
\usepackage{dcolumn}
\usepackage{bm}
\usepackage{hyperref}
\usepackage[mathlines]{lineno}
\usepackage{amsmath,bm}
\usepackage{times}
\usepackage{fouriernc}

\newcommand{\liga}{LiGaCr$_4$S$_8$}
\newcommand{\ligao}{LiGaCr$_4$O$_8$}

\begin{document}

\preprint{APS/123-QED}


\title{Negative thermal expansion and magnetoelastic coupling in the breathing pyrochlore lattice material LiGaCr$_4$S$_8$ }
\thanks{ This manuscript has been authored by UT-Battelle, LLC under Contract No. DE-AC05-00OR22725 with the U.S. Department of Energy.  The United States Government retains and the publisher, by accepting the article for publication, acknowledges that the United States Government retains a non-exclusive, paid-up, irrevocable, world-wide license to publish or reproduce the published form of this manuscript, or allow others to do so, for United States Government purposes.  The Department of Energy will provide public access to these results of federally sponsored research in accordance with the DOE Public Access Plan (http://energy.gov/downloads/doe-public-access-plan).}%

\author{G. Pokharel}
\affiliation{Department of Physics \& Astronomy, University of Tennessee, Knoxville, TN 37996, USA}
\affiliation{Neutron Scattering Division, Oak Ridge National Laboratory, Oak Ridge, TN 37831, USA}

\author{A. F. May}
\affiliation{Materials Science \& Technology Division, Oak Ridge National Laboratory, Oak Ridge, TN 37831, USA}

\author{D. S. Parker}
\affiliation{Materials Science \& Technology Division, Oak Ridge National Laboratory, Oak Ridge, TN 37831, USA}

\author{S. Calder}
\affiliation{Neutron Scattering Division, Oak Ridge National Laboratory, Oak Ridge, TN 37831, USA}

\author{G. Ehlers}
\affiliation{Neutron Technologies Division, Oak Ridge National Laboratory, Oak Ridge, TN 37831, USA} 

\author{A. Huq}
\affiliation{Neutron Scattering Division, Oak Ridge National Laboratory, Oak Ridge, TN 37831, USA}

\author{S. A. J. Kimber}
\affiliation{Neutron Scattering Division, Oak Ridge National Laboratory, Oak Ridge, TN 37831, USA}

\author{H. Suria Arachchige}
\affiliation{Department of Physics \& Astronomy, University of Tennessee, Knoxville, TN 37996, USA}
\affiliation{Neutron Scattering Division, Oak Ridge National Laboratory, Oak Ridge, TN 37831, USA}

\author{L. Poudel}
\affiliation{Department of Physics \& Astronomy, University of Tennessee, Knoxville, TN 37996, USA}
\affiliation{Neutron Scattering Division, Oak Ridge National Laboratory, Oak Ridge, TN 37831, USA}
\affiliation{Department of Materials Science \& Engineering, University of Maryland, College Park, MD 20742, USA}
\affiliation{NIST Center for Neutron Research, National Institute of Standards and Technology, Gaithersburg, Maryland 20899, USA}

\author{M. A. McGuire}
\affiliation{Materials Science \& Technology Division, Oak Ridge National Laboratory, Oak Ridge, TN 37831, USA}

\author{D. Mandrus}
\affiliation{Department of Physics \& Astronomy, University of Tennessee, Knoxville, TN 37996, USA}
\affiliation{Materials Science \& Technology Division, Oak Ridge National Laboratory, Oak Ridge, TN 37831, USA}
\affiliation{Department of Material Science \& Engineering, University of Tennessee, Knoxville, TN 37996, USA}

\author{A. D. Christianson}
\affiliation{Materials Science \& Technology Division, Oak Ridge National Laboratory, Oak Ridge, TN 37831, USA}
\affiliation{Neutron Scattering Division, Oak Ridge National Laboratory, Oak Ridge, TN 37831, USA}
\affiliation{Department of Physics \& Astronomy, University of Tennessee, Knoxville, TN 37996, USA}


\date{\today}

\begin{abstract}
The physical properties of the spinel LiGaCr$_4$S$_8$ have been studied with neutron diffraction, X-ray diffraction, magnetic susceptibility and heat capacity measurements. The neutron diffraction and synchrotron X-ray diffraction data reveal negative thermal expansion (NTE) below 111(4) K.  The magnetic susceptibility deviates from Curie-Weiss behavior with the onset of NTE.  At low temperature a broad peak in the magnetic susceptibility at 10.3(3) K is accompanied by the return of normal thermal expansion.  First principles calculations find a strong coupling between the lattice and the simulated magnetic ground state.  These results indicate strong magnetoelastic coupling in LiGaCr$_4$S$_8$.

\end{abstract}

\pacs{Valid PACS appear here}
\maketitle

\section{Introduction}

\begin{figure}
\centering
\includegraphics[width=0.9\columnwidth]{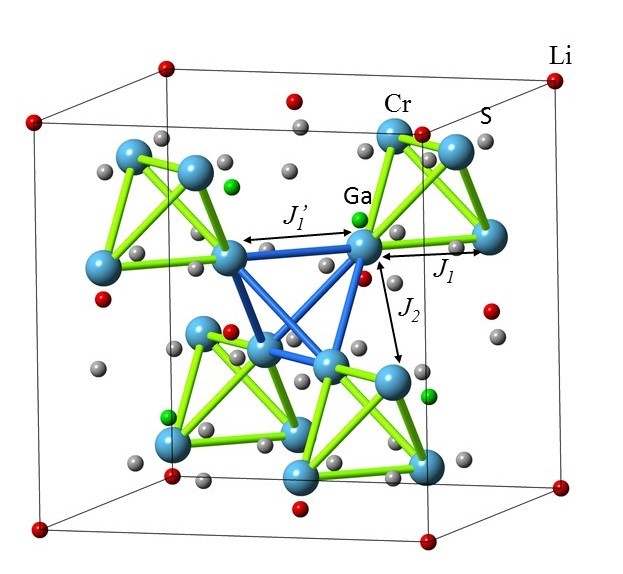}
    \caption[width= 1\textwidth]{Crystal structure of LiGaCr$_4$S$_8$ depicting the breathing pyrochlore Cr sublattice.  The exchange interactions within the small and large tetrahedra are labeled $J_1$ and $J_1'$ respectively.  $J_2$ denotes the next nearest neighbor exchange interaction.}
\label{breathing}
\end{figure}

Breathing pyrochlore lattices have emerged as an important structural motif for the realization of novel quantum phases of matter \cite{Okamoto_2013, PhysRevB.94.075146, Li2016, Neel}. A breathing pyrochlore lattice is an alteration of the pyrochlore lattice consisting of alternating large and small tetrahedra (see Fig. \ref{breathing}).  The modulation of the size can result in dramatically different exchange interactions connecting the magnetic atoms within the differently sized tetrahedra.  Recently studied examples include Ba$_3$Yb$_2$Zn$_5$O$_{11}$~\cite{kimura_2014,PhysRevB.94.075146, PhysRevB.93.220407, Rau_2016} and Li(In,Ga)Cr$_4$O$_8$\cite{Okamoto_2013}. These  materials exist in the opposite limits of the breathing pyrochlore lattice:  Ba$_3$Yb$_2$Zn$_5$O$_{11}$ is in the noninteracting limit where the individual tetrahedra are uncoupled, whereas in Li(In,Ga)Cr$_4$O$_8$ the individual tetrahedra are strongly coupled.  The investigation of model systems between these limits is a current challenge.

One starting point for the realization of a breathing pyrochlore lattice is the chromium containing chalcogenide compounds, ACr$_2$X$_4$ (X = S, Se).  These materials crystallize into a spinel structure with centrosymmetric space group $\it{Fd\bar{3}m}$ and provide a remarkably versatile playground to study the physics emerging from frustrated magnetic interactions~\cite{PhysRevB.77.115106, PhysRevB.86.104420, PhysRevLett.92.116401}. In these materials, the metal ions A$^{2+}$ occupy the tetrahedral sites forming a diamond like structure and the magnetic ions Cr$^{3+}$ occupy the octahedral sites forming a pyrochlore lattice with corner sharing Cr$_4$ tetrahedra. These materials are often multifunctional and exhibit interesting phenomena such as multiferrocity~\cite{1915197}, large magnetocapacitance/magnetoresistance~\cite{PhysRevLett.96.157202}, negative thermal expansion~\cite{ PhysRevLett.98.147203}, helical magnetism~\cite{0953-8984-28-14-146001}, strong magnetoelastic coupling \cite{PhysRevLett.97.087204, PhysRevLett.95.247204}, spin nematics~\cite{0953-8984-28-14-146001}, and spin/orbital glasses~\cite{1367-2630-6-1-191, PhysRevB.77.035207, PhysRevLett.94.027601}. This rich physical behavior is derived from the interplay of spin, charge and lattice degrees of freedom~\cite{1915197}. 

Of particular interest here, are Cr-based spinels with a non-magnetic A-site cation which often exhibit strong magnetoelastic effects \cite{PhysRevLett.94.137202, PhysRevLett.97.087204, PhysRevLett.95.247204, PhysRevLett.98.147203}. For example, the source of structural instability in chalcogenide spinels such as ZnCr$_2$S$_4$ \cite{PhysRevLett.97.087204}and ZnCr$_2$Se$_4$ \cite{PhysRevLett.98.147203} has been identified as an effect of competing ferromagnetic (FM) and antiferromagnetic (AFM) exchange interactions due to the presence of strong bond frustration. Additionally, in CdCr$_2$O$_4$ \cite{PhysRevB.87.064402}, and ZnCr$_2$Se$_4$\cite{PhysRevLett.98.147203}, the presence of strong magnetoelastic coupling drives a region of negative thermal expansion (NTE).  Hence, one of the motivations of this paper is to determine if similar physics is present in structurally-related quaternary spinels that also possess a breathing pyrochlore lattice of the magnetic species. 

A breathing pyrochlore lattice can be realized through substitution of inequivalent cations on the A-site of the aforementioned family of ternary spinels. This path has been recently explored in Cr-spinel oxides, ACr$_2$O$_4$, by substituting two inequivalent metal ions with different oxidation states such as Li$^+$ and Ga$^{3+}$/In$^{3+}$ ~\cite{Okamoto_2013, PhysRevB.93.174402}. When the substituted cations are ordered the resulting breathing pyrochlore lattice is described by non-centrosymmetric space group $\it{F\bar{4}3m}$. For example, in LiGaCr$_4$O$_8$ the cations Li$^+$ and Ga$^{3+}$ alternatively occupy the A-sites of the ACr$_2$O$_4$ spinel structure and exert an unequal local chemical pressure on the Cr$_4$ tetrahedra~\cite{ Okamoto_2013}. Thus, a breathing pyrochlore lattice forms by the arrangement of larger and smaller Cr$_4$ tetrahedra in a corner sharing network. Oxide breathing pyrochlores have been reported to order antiferromagnetically at low temperature with complex magnetostructural order \cite{Okamoto_2013, PhysRevB.91.174435} and multistage symmetry breaking~\cite{PhysRevB.93.174402}. Related S-based materials including LiGaCr$_4$S$_8$ were studied by H.L. Pinch et.al~\cite{ PINCH1970425} who reported that they order antiferromagnetically with Neel temperature between 6 K and 31 K, yet many of the basic physical properties remain unknown. Recently, additional studies of LiInCr$_4$S$_8$, LiGaCr$_4$S$_8$ and CuInCr$_4$S$_8$ have been reported \cite{Okamoto_2018} further demonstrating the rich physics of breathing pyrochlore lattice systems. 

\begin{figure}
\centering
\includegraphics[width=1\columnwidth]{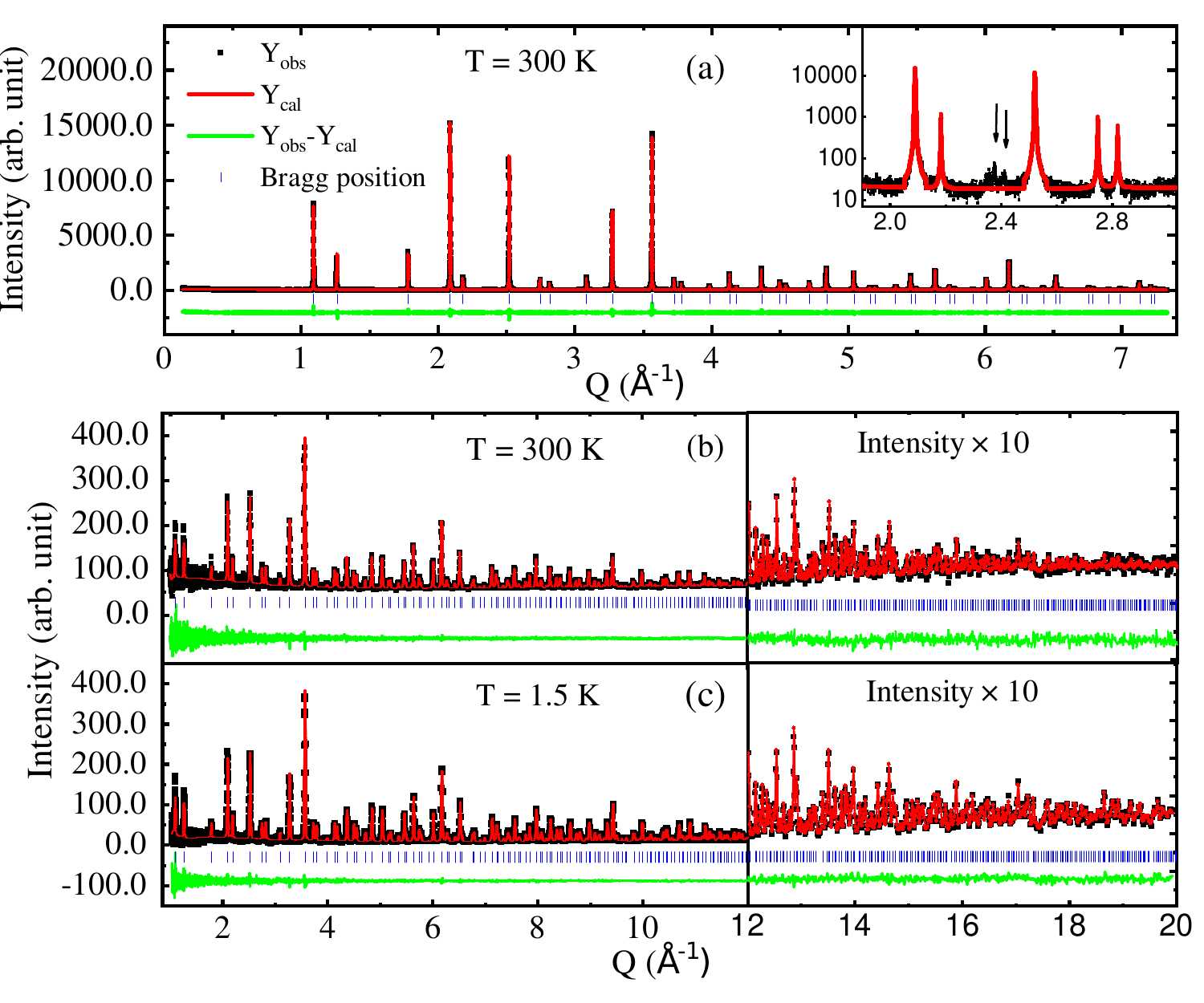}
    \caption{(a) Synchrotron X-ray diffraction data with $\lambda = 0.4146$ \AA  $ $ collected at 300 K. The inset displays a limited region of reciprocal space so that the presence of a weak Cr$_3$S$_4$ can be compared to the intensity of the main phase.  The impurity phase is indicated by arrows.  The axis labels are the same as the main panel, but note the logarithmic y-axis scale. Neutron diffraction data collected with POWGEN with  0.167 $\leq \lambda \leq 1.233$ \AA$ $ at 300 K and 1.5 K are shown in panel (b) and (c) respectively. The Rietveld refinement of the structural model described in the text with space group $\it{F\bar{4}3m}$ is indicated by the solid line through the data. }
\label{POWGEN_neutron}
\end{figure}

In this paper, we study the chalcogenide spinel LiGaCr$_4$S$_8$, where the A site of spinel structure is occupied by the two metal ions Li$^+$ and Ga$^{3+}$.  As in the oxide-based spinel LiGaCr$_4$O$_8$, the metal atoms Li and Ga are ordered and occupy the tetrahedral sites alternately with a diamond like arrangement\cite{PINCH1970425,HAEUSELER_1977} and Cr ions form a breathing pyrochlore lattice as shown in Fig. \ref{breathing}. The presence of two unequally sized Cr$_4$ tetrahedra yields a distinct nearest neighbor (NN) exchange interaction for each (here denoted by $J_1$ and $J_1'$ to allow for comparison to normal spinel counterparts). One means of understanding the degree to which the different sizes of the tetrahedra may modify the physical behavior is through the breathing ratio ($\it{B_r}$), defined as the ratio of the Cr-Cr bond length within the larger and smaller Cr$_4$ tetrahedra, $\it{B_r}$ = d'/d $\ge$ 1.  For the A-site ordered quaternary spinels, the breathing ratio can be determined from the $x$-coordinate of the Cr$^{3+}$ ion via $\it{B_r}$ = $|(2x-1)/(2x-0.5)|$.  Importantly, this shows that the breathing ratio is independent of the lattice parameter. 

\begin{figure}
\centering
 \includegraphics[width=0.95\columnwidth]{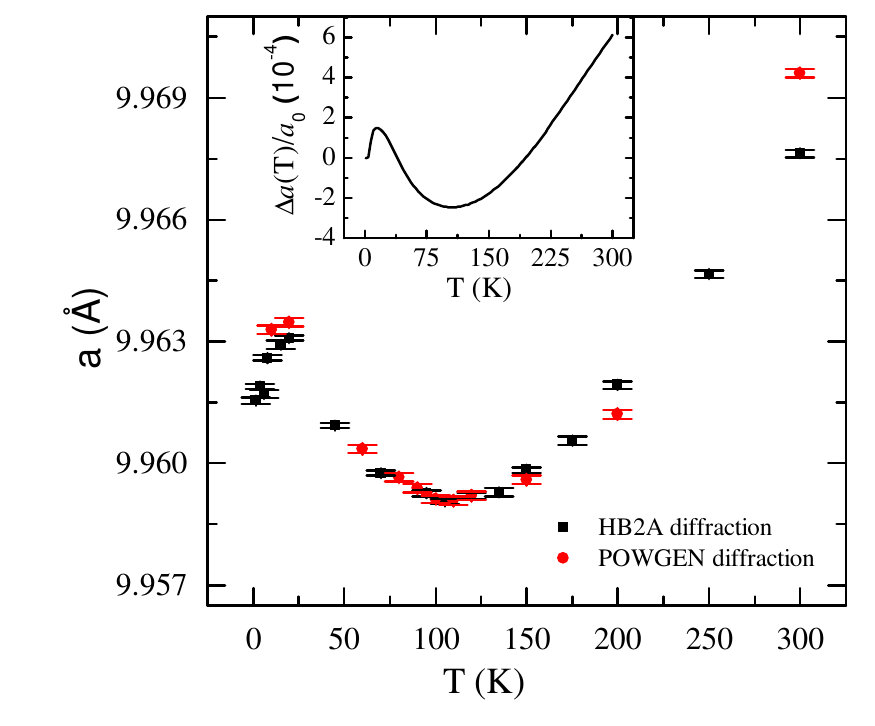}
    \caption[width= 0.35\textwidth]{Lattice parameter of LiGaCr$_4$S$_8$ as a function of temperature obtained by neutron diffraction. Negative thermal expansion is observed below 111(4) K. The inset shows the estimated temperature dependence of $\Delta a(T)/a_0$ determined as explained in the text.} 
\label{thermalexp}
\end{figure}

We use neutron and X-ray diffraction to study the structural properties of polycrystalline LiGaCr$_4$S$_8$. We find a $\it{B_r}$ of 1.077(2) which represents a modest increase over the value of 1.035(1) for LiGaCr$_4$O$_8$ \cite{Okamoto_2013}.  Below 111(4) K, we find a region of NTE which extends down to 10 K. This behavior is similar to that observed in the spinel ZnCr$_2$Se$_4$\cite{PhysRevLett.98.147203} however a similar region of NTE has not been observed in related oxide-based breathing pryochlores such as \ligao{}. The changes in lattice expansion are accompanied by the departure from Curie-Weiss behavior of the magnetic susceptibility and a peak in the magnetic susceptibility and specific heat at 10.3(3) K.  This indicates that magnetic and lattice degrees of freedom are strongly coupled in LiGaCr$_4$S$_8$.  Additionally, first principles calculations presented here also find strong coupling between magnetism and the lattice.

\section{Experimental Details}

Polycrystalline samples of LiGaCr$_4$S$_8$ were synthesized by solid state reaction. Stoichiometric amounts of Ga (99.999\%), Li$_2$S (99.9\%), Cr (99.95\%), S (99.9995\%), purchased from Alfa Aesar, were ground together inside a glove box and then pressed into a 0.5 inch diameter pellet. The resulting pellet was heated to 1175 K for 3 days. This process was repeated until laboratory X-ray diffraction patterns indicated a phase pure sample.

Synchrotron X-ray diffraction measurements were performed at 11-BM the Advanced Photon Source (APS) at Argonne National Laboratory (ANL) using X-rays of wavelength $\lambda$ =  0.4146 \AA. For the measurements, a finely ground polycrystalline sample was packed inside a Kapton tube with a diameter of 0.8 mm, which was mounted on the cold finger of an Oxford helium cryostat. The sample was spun at the frequency of 50 Hz to ensure proper powder randomization. The Rietveld refinement packages FullProf\cite{bx} and GSAS/EXPGui \cite{J_Appl_Crystl} were used to refine the crystal structure against the X-ray diffraction data. 

Neutron diffraction data were collected using the HB-2A powder diffractometer at the High Flux Isotope Reactor at Oak Ridge National Laboratory (ORNL). Measurements were made at HB-2A in the temperature range 1.5 K to 300 K using incident neutrons with wavelength, $\lambda$ = 1.5396 \AA. Additional neutron diffraction measurements were made with POWGEN at the Spallation Neutron Source, ORNL using a band of incident neutrons with 0.167 $\leq \lambda \leq 1.233$ \AA.  The Rietveld refinement packages FullProf\cite{bx} and GSAS/EXPGui \cite{J_Appl_Crystl} were used to refine the crystal structure against the neutron diffraction data. 

\begin{figure}
\centering
 \includegraphics[width=0.85\columnwidth]{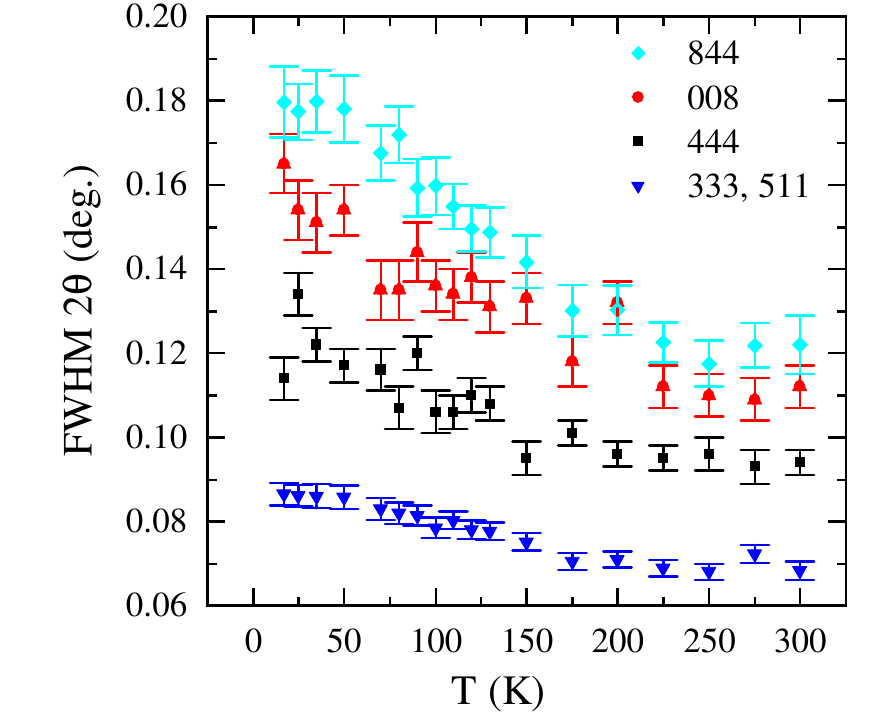}
    \caption[width= 0.35\textwidth]{Temperature dependence of peak width (Full Width Half Maximum of a standard Lorentzian function) of selected diffraction peaks determined from the laboratory X-ray diffraction data. All peaks broaden with cooling at low temperature.}. 
\label{peakwidth}
\end{figure}

Field cooled (FC) and zero field cooled (ZFC) dc susceptibility measurements with applied magnetic fields $H$ ranging from $\mu_O H$ of 0.01 to 5 T in the temperature range 2 K to 300 K were carried out using a Quantum Design magnetic property measurement system (MPMS). AC magnetic susceptibility measurements were performed in a Quantum Design physical property measurement system (PPMS).  These measurements were performed with a static field of $H_{dc}$ = 0, using an ac amplitude of $H_{ac}$ = 5 Oe.  The frequency $f$ dependence of the in-phase component $\chi \prime$ was examined by performing measurements as a function of $f$.  Data were collected upon cooling from 30 to 2 K. Heat capacity was measured using a Quantum Design PPMS in the temperature range 2 K to 300 K. Measurement of the electrical resistivity  was carried out on a sintered sample with a Quantum design PPMS using the four-point measurement technique with gold wires attached to the sample using silver paint.

\section{Results}   

\subsection{Neutron Diffraction and X-ray Diffraction}

We begin with a discussion of the crystallography and lattice behavior of  LiGaCr$_4$S$_8$. Neutron and synchrotron X-ray diffraction patterns are shown in Fig. \ref{POWGEN_neutron}. Unlike the diffraction pattern corresponding to a normal spinel structure with space group $\it{Fd\bar{3}m}$ (No. 227), additional reflections, such as 002 are observed as expected for space group $\it{F\bar{4}3m}$ (No. 216) \cite{PINCH1970425,HAEUSELER_1977}. The space group $\it{F\bar{4}3m}$ is a subgroup of space group $\it{Fd\bar{3}m}$. The number of symmetry operations are reduced by half, including the loss of inversion symmetry in $\it{F\bar{4}3m}$ compared to the parent space group.  The diffraction patterns are consistent with those of the oxides of related breathing pyrochlore lattice materials LiGaCr$_4$O$_8$ and LiInCr$_4$O$_8$ ~\cite{Okamoto_2013} and studies of \liga{} in Ref. \cite{Okamoto_2018}.  We find only weak impurity phases: Cr$_3$S$_4$ and an even less significant unidentified phase (see Inset of Fig. \ref{POWGEN_neutron}(a)). We estimate the impurity phase fraction to be less than one percent of the main phase based upon taking ratios of Bragg peak intensities. On the other hand Ref. \onlinecite{Okamoto_2018} finds impurity phases of less than a few percent of Cr$_2$S$_3$ and an additional unknown phase.

Rietveld refinement of the structural model against the X-ray data, as well as the neutron diffraction data discussed below, confirms that the metal ions Li$^{+}$ and Ga$^{3+}$ occupy the A-sites of the spinel structure with occupancies close to the ideal values of equivalent sites in the standard spinel structure (Fd$\bar{3}$m). Initial refinements assumed these 4a and 4d sites were fully occupied by Li and Ga respectively.  Better agreement with the data was obtained by allowing Li and Ga site interchange, and this approach was informed by the complementary nature of neutron and x-ray diffraction data.  The best refinements indicate the Ga site is fully occupied by Ga and in final refinements the occupancy of this site was fixed. The Li site was found to have some Ga, however, and refinements suggest a $\approx$5$\%$ deficiency of Li in the samples studied here. The refined composition of Li$_{0.956}$Ga$_{1.043}$Cr$_4$S$_8$ is obtained from a  combined refinement of both the neutron and synchrotron x-ray data, and the additional refinement results are presented in Table \ref{parm_nd}.

Initial x-ray diffraction data indicated a region of NTE.  To examine this behavior over a broader temperature range, neutron powder diffraction measurements were performed and the lattice parameter as a function of temperature was extracted. The variation of lattice parameter with temperature is shown in Fig. \ref{thermalexp}.  Cooling from room temperature, the lattice parameter decreases until 111(4) K. Further cooling results in lattice expansion (NTE) until $\approx$10 K where the NTE terminates. Parameters obtained by a combined refinement of the structural model to the synchrotron X-ray and POWGEN neutron diffraction data are shown in Table \ref{parm_nd} (a). The refined parameters at 10 K obtained from POWGEN neutron diffraction are also shown in Table \ref{parm_nd} (b). No direct evidence for a distortion from cubic symmetry was detected in any of our diffraction data, including the high-resolution synchrotron data. However, below $\sim$200 K a temperature-dependent broadening of the Bragg peaks was observed in the x-ray diffraction data upon cooling (see Fig.\ref{peakwidth}), likely indicating the development of microstrain.

This behavior can be compared to other NTE materials such as ZnCr$_2$Se$_4$\cite{PhysRevLett.98.147203} by comparing the relative change in lattice parameter as a function of temperature, $\Delta a(T)/a_0$. Here $\Delta a(T)/a_0$ of LiGaCr$_4$S$_8$ has been estimated using a cubic spline interpolation of the data. For this estimation, the lattice parameter at 1.5 K is taken as the standard initial value, $a_0$. $\Delta a(T)/a_0$ increases up to $\sim$12 K.  $\Delta a(T)/a_0$ exhibits NTE in the temperature range 12(2) - 111(4) K.   

\begin{figure}[tp]
\centering
 \includegraphics[width= 0.45\textwidth]{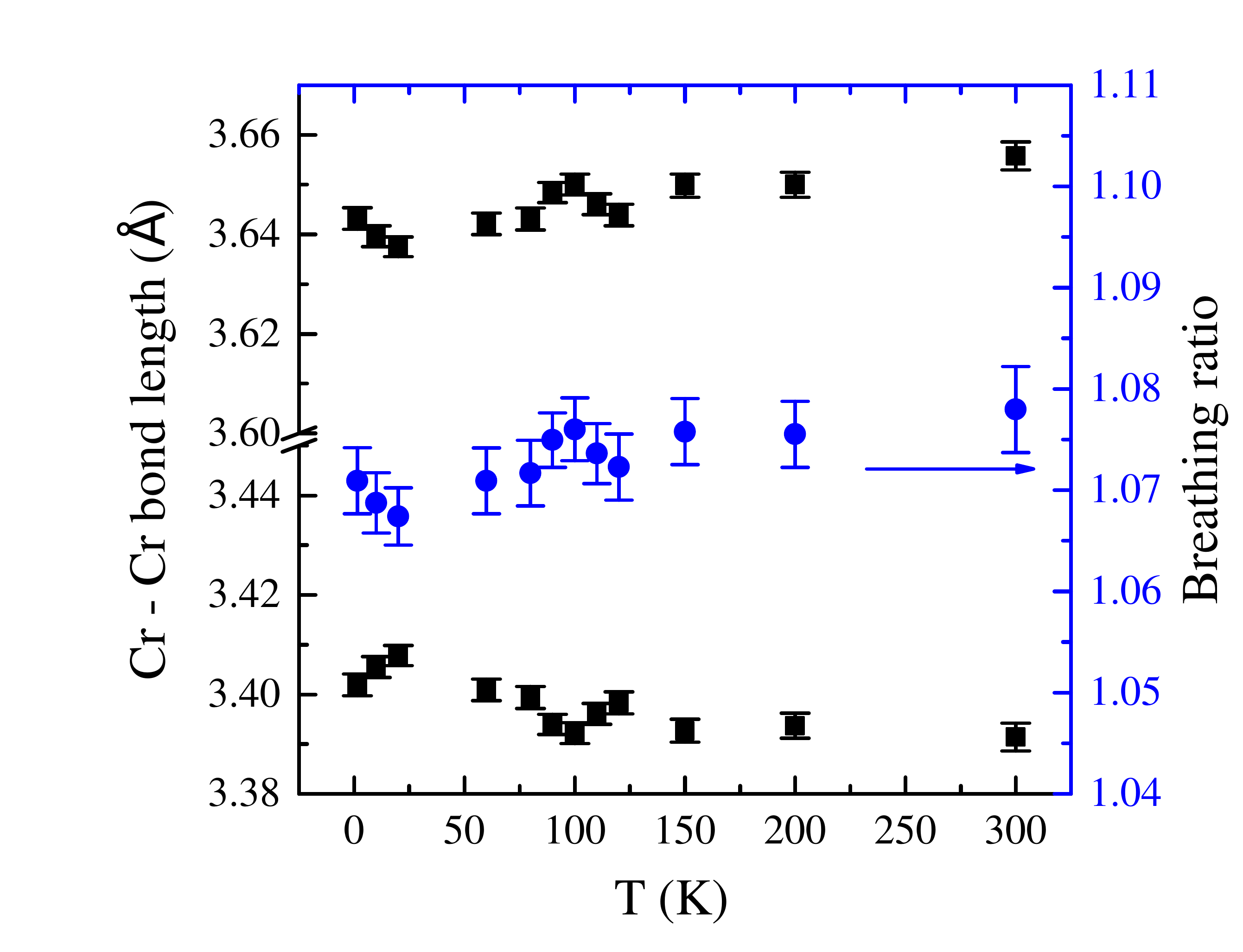}
    \caption[width= 0.95\columnwidth]{Temperature dependent Cr-Cr bond lengths in the small and large Cr$_4$-tetrahedra (filled squares, left scale) and $B_r$ (filled circles, right scale).
     These bond lengths and $B_r$s are obtained from the neutron diffraction data collected with POWGEN. }
    \label{Bond_length}
\end{figure}

\begin{figure}[tp]
\centering
 \includegraphics[width= 0.45\textwidth]{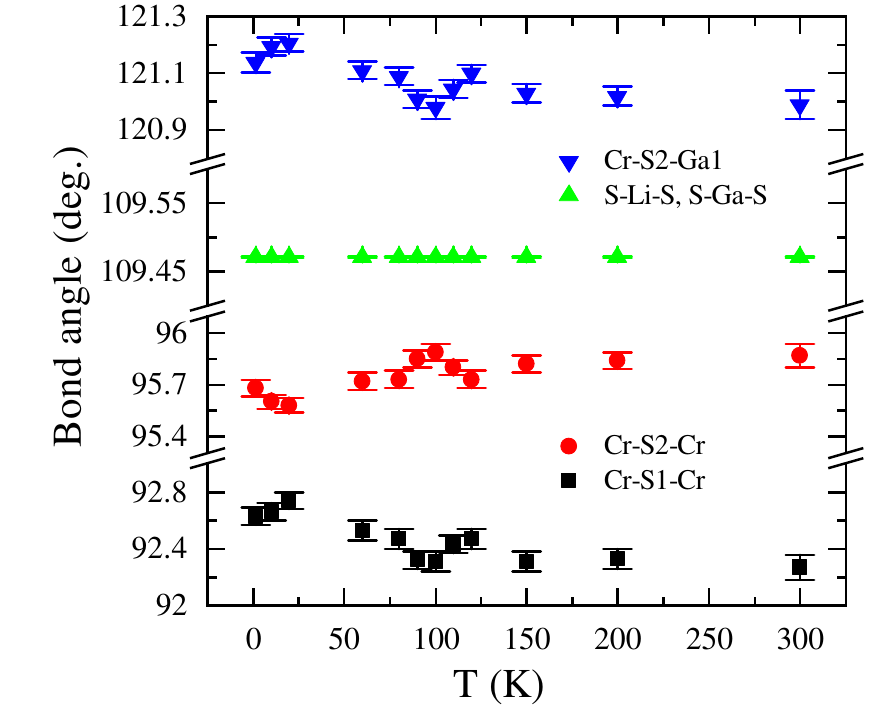}
    \caption[width= 0.35\textwidth]{Variation of selected bond angles with temperature in LiGaCr$_4$S$_8$. Cr-S-Cr bonds are close to 90$^{\circ}$ and are responsible for the NN superexchange interaction. S-Li-S and S-Ga-S bonds are insensitive to the change in temperature whereas other bonds show a weak anomaly with the onset of NTE.}
    
    \label{angle}
\end{figure}

The onset of NTE is reflected in the internal degrees of freedom of the unit cell.  The effect is evident in bond lengths and bond angles involving Cr.  The two Cr-Cr bond lengths show a weak anomaly near the onset of negative thermal expansion (Fig. \ref{Bond_length}). Changes in the two Cr-S-Cr bond angles are also observed (Fig. \ref{angle}).  Both the Cr-Cr distances and Cr-S-Cr bond angles are expected to play an important role in the magnetic exchange interactions in LiGaCr$_4$S$_8$ and the observations of changes in these quantities appear correlated with the changes in magnetic behavior that are discussed in following sections.

\begin{table}
\centering
\caption{(a)  Parameters of LiGaCr$_4$S$_8$ obtained from combined refinement of the structural model to the synchrotron X-ray and POWGEN neutron diffraction data.}
\begin{tabular}{c c c c c}
\multicolumn{5}{c}{T = 300 K, $\it{R_p}$= 6.9 \%, $\it{R_{wp}}$= 4.2 \%} \\ 
\multicolumn{5}{c}{$\chi^2$ = 1.4, $\it{a}$ = 9.9696(1) \AA} \\

\hline
atom & $\it{x}$ = $\it{y}$ = $\it{z}$ & $\it{B_{d}}$ (\AA$^2$) & $\it{B_{nd}}$ (\AA$^2$) & occupancy \\ [0.5ex]
\hline\hline
S$_1$($\it{16e}$) & 0.1342(1) & 0.69(2) & 0.08(2) & 1 \\ 
S$_2$($\it{16e}$) & 0.616(1) & 0.61(2) & -0.06(2) & 1 \\
Cr($\it{16e}$) & 0.370(1) & 0.60(1) & 0.00(2) & 1 \\
Ga$_1$($\it{4d)}$ & 0.75 & 0.63(1) & 0 & 1 \\
Li($\it{4a}$) & 0 & 1.24(7) & 0 & 0.953(1) \\
Ga$_2$($\it{4a}$) & 0 & 1.24(7) & 0 & 0.046(1)  \\ [1ex]
\hline
\end{tabular}
\bigskip

\begin{tabular}{c c c c c}
\multicolumn{5}{c}{(b) Refined parameters of LiGaCr$_4$S$_8$ obtained from}\\
\multicolumn{5}{c}{POWGEN neutron diffraction.}\\
\multicolumn{5}{c}{T = 10 K, $\it{R_p}$= 5.8 \%, $\it{R_{wp}}$= 1.9 \% } \\ 
\multicolumn{5}{c}{$\chi^2$ = 1.6, $\it{a}$ = 9.9633(3) \AA} \\
\hline
atom & $\it{x}$ = $\it{y}$ = $\it{z}$ & $\it{B_{d}}$ (\AA$^2$) & $\it{B_{nd}}$ (\AA$^2$) & occupancy \\ [0.5ex]
\hline\hline
S$_1$($\it{16e}$) & 0.1345(1) & 0.33(2) & 0.02(1) & 1 \\ 
S$_2$($\it{16e}$) & 0.6167(1) & 0.25(1) & -0.02(1) & 1 \\
Cr($\it{16e}$) & 0.3708(1) & 0.29(1) & 0.04(1) & 1 \\
Ga$_1$($\it{4d}$) & 0.75 & 0.23(2) & 0 & 1 \\
Li($\it{4a}$) & 0 & 1.10(8) & 0 & 0.953 \\
Ga$_2$($\it{4a}$) & 0 & 1.10(8) & 0 & 0.046  \\[1ex]
\hline
\multicolumn{5}{c}{$\it{B_{d}}$ = Anisotropic diagonal thermal parameter} \\ 
\multicolumn{5}{c}{$\it{B_{nd}}$ = Anisotropic non-diagonal thermal parameter} \\

\end{tabular}
\label{parm_nd}
\end{table}

\begin{figure}[tp]
\centering
 \includegraphics[width= 0.49\textwidth]{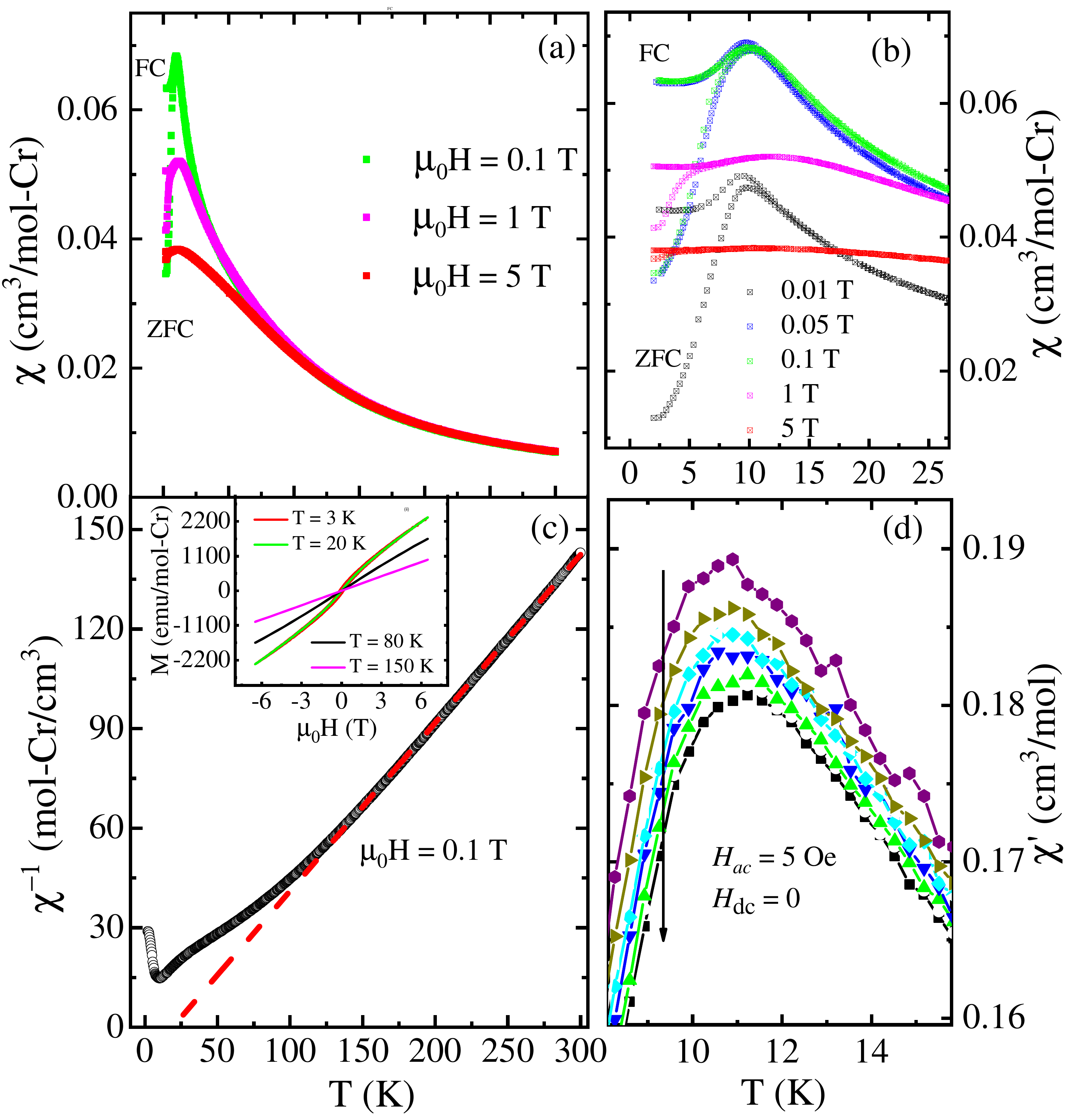}
    \caption[width= 0.35\textwidth]{Magnetic susceptibility measurements of LiGaCr$_4$S$_8$. (a, b) Show the temperature dependent dc susceptibility, $\chi$ and (c) shows the temperature dependent inverse dc susceptibility, $1/\chi$ of LiGaCr$_4$S$_8$. (d) Displays the in-phase part of the ac susceptibility, $\chi'$ around the magnetic transition, measured with $\it{H_{ac}}$ = 5 Oe with frequencies 29 Hz, 127 Hz, 1129 Hz, 2336 Hz, 4832 Hz and 10000 Hz. The arrow in (d) indicates the direction of increasing applied frequency. The inset of (c) displays $M(H)$ loops measured at selected temperatures. A peak in the susceptibility occurs at 10.3(3) K and deviation from Curie-Weiss behavior is observed below 110 K. Small shift in the peak of $\chi'$ towards higher temperature is noticed with the increase in frequency. Weak hysteresis in the $M$ vs. $H$ data is observed at 3 K. }
\label{susc}
\end{figure}

As described in the Introduction, the difference between the two Cr-Cr bond lengths is driven by the inequivalent ionic radii (59 pm and 47 pm\cite{shannon_1976}) of the Li$^{1+}$ and Ga$^{1+}$ ions respectively. This difference in bond lengths is quantified by $\it{B_r}$. At 300 K the larger and smaller Cr-Cr bond lengths d' and d extracted from the neutron diffraction data are found to be 3.655(2) \AA and 3.394(2) \AA yielding a breathing ratio $\it{B_r}$ = 1.077(2). The value of $\it{B_r}$ as a function of temperature is shown in Fig. \ref{Bond_length}. The value $\it{B_r}$ in LiGaCr$_4$S$_8$ is somewhat larger than the values of 1.035(1) and 1.051(1) for LiGaCr$_4$O$_8$ and CuInCr$_4$O$_8$ respectively \cite{Okamoto_2013}. However, the breathing ratio is considerably less than in Ba$_3$Yb$_2$Zn$_5$O$_{11}$, where $\it{B_r}$=1.90(2) is observed \cite{kimura_2014}. The value of $\it{B_r}$ and the thermodynamic measurements presented in the following sections places \liga{} in the interacting limit of the breathing pyrochlore lattice.

\subsection{Magnetic Properties}\label{magnetic properties}

To study the magnetic properties of LiGaCr$_4$S$_8$, temperature-dependent magnetization $M$ measurements were performed under applied fields $\mu_O H$ ranging from 0.01 to 5 T, and the results are summarized in Fig. \ref{susc}. A cusp in the susceptibility $\chi = M/H$ is evident at 10.3(3) K (Fig. \ref{susc} (a)), indicating a magnetic transition. The zero field cooled and field cooled curves bifurcate below the cusp, and the temperature where bifurcation ceases is inversely proportional to applied field (Fig. \ref{susc}(b)).  This behavior is commonly observed in systems with glassy spin dynamics, and such spin freezing transitions are often observed in the spinel family due to the strong frustration and competing interactions.  The weak hysteresis observed in the field-dependence of the magnetization is also consistent with glassy-dynamics below $\approx$10 K (inset, Fig. \ref{susc}(c)).  To verify that the cusp near 10 K is associated with a transition involving glassy spin dynamics, we performed a time-dependent measurement below 10 K and observed that the remanent moment does indeed decrease with increasing time.  In addition, we have performed ac susceptibility measurements which show a weak frequency dependence of the peak in the in-phase part of the susceptibility, $\chi \prime$ (see Fig. \ref{susc}(d)).   Thus, the cusp in $\chi$ near 10 K is likely associated with some type of spin freezing transition, the nature of which is still under investigation.

From 150 to 300 K, $1/\chi$ is linear in $T$ (Fig. \ref{susc}(c)) and thus the  Curie-Weiss law provides an excellent description of the data.  However, at temperatures below 125 K, the susceptibility does not increase as fast as expected from Curie-Weiss behavior, and this may reflect the increasing importance of antiferromagnetic correlations upon cooling.  Importantly, the departure from Curie-Weiss behavior appears to coincide with the onset of NTE. Fitting the data above 150 K to the Curie-Weiss law, $\chi=C/(T-\Theta_{CW})$, produces a Curie constant $C$ that yields an effective moment of $\mu_{eff}$ = 3.96(1) $\mu_B$/Cr, and a Weiss Temperature $\Theta_{CW}$= 19.5(1) K. The value of $\mu_{eff}$ is slightly larger than the theoretical value of 3.86 $\mu_B$ for S=3/2. The positive $\Theta_{CW}$ indicates  ferromagnetic correlations dominate in the high-$T$ regime. We note that including a diamagnetic contribution in the fitting procedure still results in a positive $\Theta_{CW}$ of $\sim$10 K. In their studies of \liga{}, Ref. \onlinecite{Okamoto_2018} obtain a negative $\Theta_{CW}$ of -20 K.  The reason for the difference with the results presented here is not readily apparent but may be due to different impurity levels in the samples or differing amounts of Li-deficiency.  The results presented here along with the presence of the short range or glassy magnetic order below 10.3 K, indicate substantial competition between AFM and FM interactions in \liga{}.

\begin{figure}
\centering
 \includegraphics[width= 0.5\textwidth]{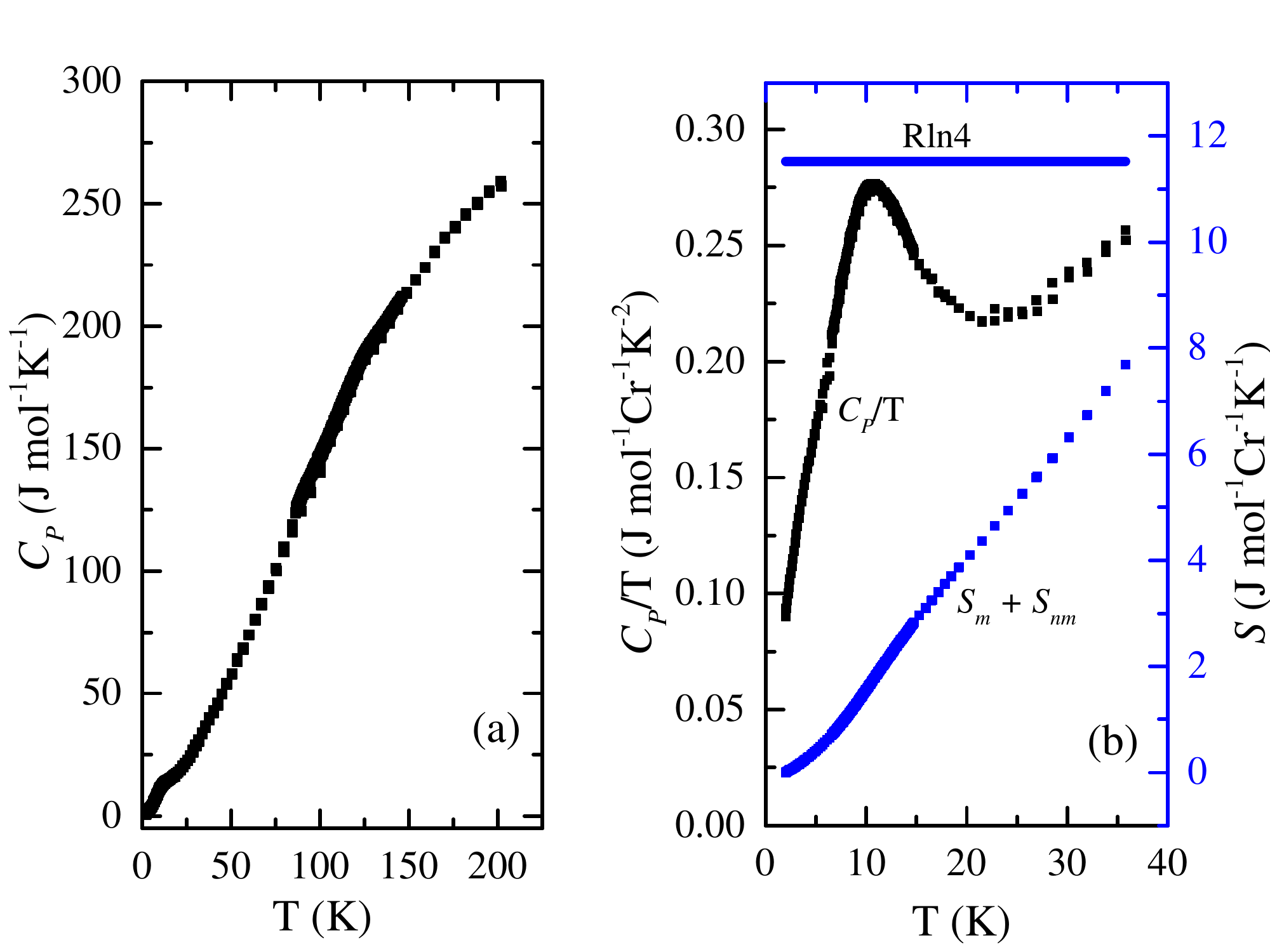}
    \caption[width= 0.35\textwidth]{Temperature dependent heat capacity, $\it{C_P}$, of \liga{} at zero magnetic field. (a) Displays the $\it{C_P}$ from 2 to 200 K.  (b) Shows $\it{C_P}$/T and the estimated entropy, S, at low temperatures. The theoretical value of magnetic entropy for spin 3/2 system is indicated by the horizontal line in (b). The change in slope of $\it{C_P}$ around at 120 K and a broad hump at 10.9(4) K look coincident with the bounding temperatures of NTE.
    }
\label{heatcap}
\end{figure}

\subsection{Heat Capacity}

The temperature dependent specific heat, $\it{C_P}$, of LiGaCr$_4$S$_8$ is displayed in Fig. \ref{heatcap}.  A broad hump in $\it{C_P}$ occurs with a maximum near 10.9(4) K.  A very weak anomaly is also observed around 120 K.  While a feature is visible, and occurs near the change in thermal expansion behavior, it is too weak to provide any additional information.  This observation is consistent with the inability of our synchrotron diffraction data to detect a structural transition near the onset of the negative thermal expansion.

The anomaly in the specific heat capacity that is associated with the magnetic transition at low $T$ is best observed in a plot of $\it{C_P}$/T vs. T (Fig. \ref{heatcap}(b)). From these data, the entropy as a function of temperature is obtained by integrating $\it{C_P}$/T dT (\ref{heatcap}(b)) starting at $T$ = 2 K.  Note that this estimate of the entropy includes all contributions to $\it{C_P}$ and represents an upper limit to the magnetic contribution to the entropy, S$_{mag}$ (no baseline is utilized as a non-magnetic analogue is not present).  In this context, it is noteworthy that the entropy determined is dramatically reduced from the value of RLn(4) expected from a magnetic order-disorder transition involving S=3/2 Cr$^{3+}$. This could be caused by the presence of residual entropy due to the glassy phase.  However, a substantial portion of the entropy may be lost at higher temperatures, corresponding to the likely interplay of magnetic and lattice degrees of freedom with an onset near 111 K. 

\begin{figure}[tp]
\centering
 \includegraphics[width= 0.4\textwidth]{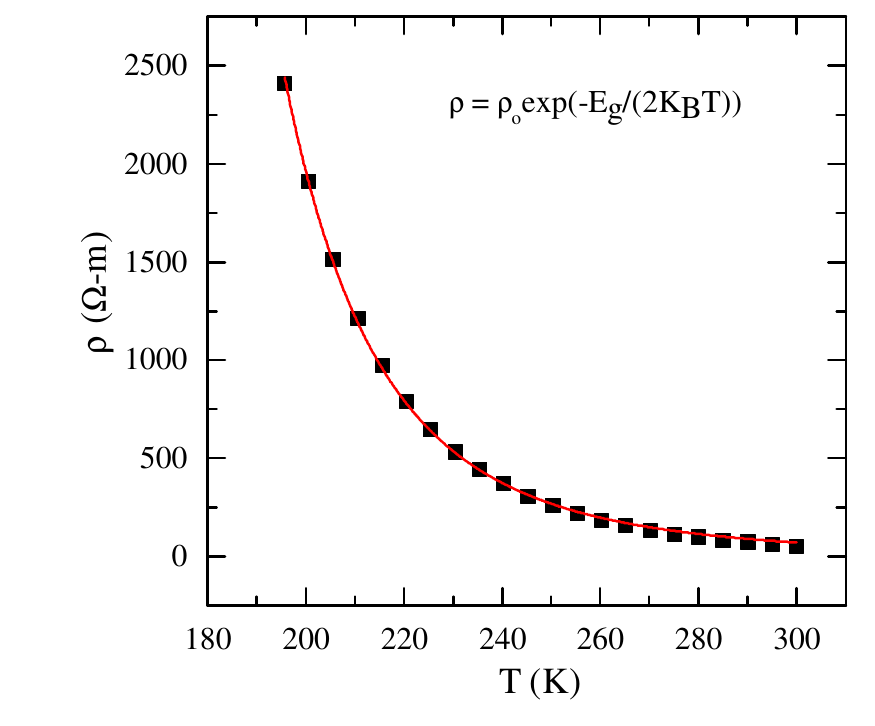}
    \caption{Temperature dependent resistivity, $\rho$ of LiGaCr$_4$S$_8$. The fitting function is described in the text. }
\label{rho}
\end{figure}

\subsection{Electrical Resistivity}

The electrical resistivity, $\rho$ of LiGaCr$_4$S$_8$ measured in the temperature range 195 to 300 K is shown in Fig \ref{rho}. Below 195 K, the resistance was too large for the instruments utilized. From 300 K, the resistivity increases exponentially with decreasing temperature indicating typical semiconducting behavior. To estimate the band gap, E$_g$, the equation $\rho(T)$ = $\rho_0$ exp(-E$_g$/2k$_B$T) is fitted to the resistivity data, where k$_B$ is the Boltzmann constant. This yields E$_g$=0.37(3) eV.   

\section{First Principles Calculations - Magnetoelastic Coupling}\label{firstprinc}

To understand the observed behavior, first principles calculations of the structure and magnetic behavior were performed, using the all-electron planewave density functional theory (DFT) code WIEN2K\cite{wien}.  The generalized gradient approximation of Perdew, Burke and Ernzerhof \cite{perdew} was employed, with an $RK_{max}$ of 7.0. Here $RK_{max}$ is the product of the smallest muffin-tin radius (that for S) and the largest planewave expansion wavevector. We assume an ordered structure with crystallographically separate Li and Ga sites, generally consistent with the experimental refinements showing only a few percent of Li and Ga mixing on these sites. 

The experimental results suggest a strong coupling of magnetism and structure in LiGaCr$_4$S$_8$, in particular with the inverse susceptibility first deviating from linear behavior at nearly the same temperature as the onset of the NTE. While we do not directly address the NTE here, these calculations also find evidence for magnetoelastic coupling.

First principles calculations were performed for two configurations: a non-magnetic configuration, and a simple ferromagnetic (FM) configuration.  While the presence of chromium, the highly electronegative sulfur, and the complex geometric frustration may argue against a simple ferromagnetic ground state, for the purposes of discussing magnetoelastic coupling this structure appears to be sufficient.  For both calculations, the internal coordinates were optimized, and a cubic lattice parameter of $a$ = 9.9675 \AA was utilized.

The optimized FM calculation finds a Cr coordinate of 0.3791, which leads to a nearest-neighbor Cr-Cr distance of 3.408 \AA, in excellent agreement (within 0.5 percent) with the experimental value of 3.390 \AA. However, the optimized non-magnetic configuration finds a Cr coordinate of 0.3956, and thereby a nearest-neighbor Cr-Cr distance of just 2.86 \AA, which
differs from the experiment by more than 0.5 \AA.  This is more than an order of magnitude beyond any possible error associated with the inherent approximations in DFT. Rather, it is suggestive that even at 300 K that the magnetism is affecting and possibly determining the structure.  While it has long been known that signatures of magnetism often persist well above the ordering point, a factor of 30 here is unusual. This is likely due to the effective suppression of the ordering point by geometric and magnetic frustration, as well as the strength of the magnetic interactions involved. We find the ferromagnetic state (with a Cr moment of approximately 3 $\mu_B$) to fall some $\sim$ 500 meV per Cr below the non-magnetic state.  

This is a relatively large energy compared with the 300 K thermal energy $k_{B}T$ of just 25 meV, and speaks to the likelihood of ``disordered local moments" persisting up to and even beyond room temperature.  By this we mean that while long-range magnetic order is absent above the ordering temperature, the individual Cr atoms likely carry a moment of approximately 3 $\mu_B$, but these moments are essentially randomly oriented in spatial direction, with relatively little moment direction correlation between neighboring Cr atoms.  Detailed discussions of this scenario applied to iron can be found in Ref. \cite{stocks}.  In this context the observed anomalous structural behavior, such as the NTE, can be considered to arise from the detailed temperature dependence of the ordering energy (via competing interactions) and its interactions with the lattice.

\section{Discussion} 

As described above, the magnetic properties of \liga{} are strongly reflected in the behavior of the lattice.  NTE appears when the susceptibility deviates from Curie-Weiss behavior. The lattice continues to expand with decreasing temperature until the magnetic transition at 10 K after which the lattice contracts to the lowest temperatures measured (1.5 K).  These observations along with the first principles calculations presented in Sec. \ref{firstprinc} are compelling evidence that magnetoelastic coupling is strong in \liga{}.   As noted above neither the neutron diffraction data nor the synchrotron X-ray data provide direct evidence of a departure from cubic symmetry.  This appears to be one of the distinctions between \liga{} and LiGaCr$_4$O$_8$, where two different phases either tetragonal or orthorhombic along with a cubic phase are reported below the ordering temperature of 12 K~\cite{PhysRevB.91.174435, PhysRevB.94.064420}.
 
The existence of NTE in magnetically frustrated Cr-spinel compounds has been explained as a consequence of magnetoelastic coupling~\cite{PhysRevLett.98.147203, 0953-8984-28-18-18LT01}. This mechanism relies on  competing AFM and FM exchange interactions\cite{hemberger_2006}. This situation is likely realized in \liga{} as the competition between antiferromagnetic and ferromagnetic exchange interactions is evident in the magnetic properties.  The bond angle for the Cr-S-Cr nearest neighbor superexchange paths are close to 90$^{\circ}$ (see Fig. \ref{angle}) which can yield a ferromagnetic exchange interaction. This is in accord with the fitting of $\chi$ with the Curie-Weiss law in the paramagnetic region from 150 to 300 K, which yields a positive $\Theta_{CW}$=19.5 K.  This indicates that ferromagnetic interactions dominate the magnetic properties at high temperature. This is another distinction between the oxide counterparts of \liga{} where antiferromagnetic correlations dominate at high temperature as evidenced by the reported $\Theta_{CW}$ values of -658.8 K and -331.9 K for LiGaCr$_4$O$_8$ and LiInCr$_4$O$_8$ respectively~\cite{Okamoto_2013}. 

At lower temperatures, antiferromagnetic correlations become important in \liga{}.  In particular, the magnetic phase transition at $\sim$10 K exhibits glassy behavior that would only be expected with the presence of antiferromagnetic couplings, and the deviation from Curie-Weiss behavior near the onset of NTE can be considered to have AFM-like character (relative decrease in $\chi$).  We note that the glassy behavior could be promoted by the deviation from perfect LiGaCr$_4$S$_8$ stoichiometry, which was demonstrated via diffraction data that revealed our samples possess excess Ga.

There are at least two possible sources for antiferromagnetic exchange in Cr-spinel compounds\cite{PhysRev.151.367,PhysRevB.77.115106}.  One is direct Cr-Cr exchange, which can be relevant in oxides where significantly smaller Cr-Cr distances are found.  In \liga{} the minimum Cr-Cr distance is 3.39 $\AA$ and thus direct exchange is not likely to be the dominate source of antiferromagnetic interactions.  A second and likely more relevant set of interactions are between 2nd and 3rd nearest neighbors \cite{Takagi2011, PhysRevB.77.115106}. In contrast to $J_1$ and $J_1'$ where there are 3 NN, there are 12 NNN each for $J_2$ and $J_3$.  Inelastic neutron scattering experiments could be useful to provide greater details to the various terms in the spin Hamiltonian.

Finally, we comment on the nature of the low temperature phase. Glassy dynamics likely plays an important role in the low temperature physical properties as evidenced from the following experimental observations:  The bifurcation of the field cooled and zero field cooled susceptibility and the weak frequency dependence of the ac susceptibility. Likewise, the observed hysteresis in the field-dependence of the magnetization is likely associated with time-dependent relaxation of the moments.  Additionally, the anomaly in the heat capacity is rather broad and analysis of this data result in a small fraction of the entropy expected for a $S=3/2$ system.  This behavior may be explained as a consequence of the chemical disorder induced by the deficiency of Li and associated occupation of the Li site by Ga (see Tables \ref{parm_nd}).  Such disorder may be particularly important for exchange interactions beyond NN. Additional efforts to understand the low temperature phase are underway.

\section{Conclusion}

In summary, we find that the chalcogenide spinel LiGaCr$_4$S$_8$ forms a breathing pyrochlore lattice with the tetrahedral A sites alternately occupied by metal ions Li$^+$ and Ga$^{3+}$.  Negative thermal expansion is observed from 10 to 110 K and appears along with the deviation from Curie-Weiss behavior of the magnetic susceptibility. At lower temperatures a magnetic transition to a phase with slow dynamics occurs at 10.3(3) K and is accompanied by the return to normal thermal expansion. Together these experimental observations along with first principles calculations point to strong magnetoelastic coupling in \liga{}.

\begin{acknowledgments}
We thank C. Batista for useful discussions and S. Lapidus for help with the x-ray diffraction measurements.  ADC, AFM, MAM and DSP were supported by the U.S. Department of Energy, Office of Science, Basic Energy Sciences, Materials Sciences and Engineering Division. GP and DM acknowledge support from Gordon and Betty Moore Foundation's EPiQS Initiative through Grant GBMF4416. This research used resources at the Spallation Neutron Source and the High Flux Isotope Reactor, a Department of Energy (DOE) Office of Science User Facility operated by Oak Ridge National Laboratory (ORNL). Use of the Advanced Photon Source at Argonne National Laboratory was supported by the U. S. Department of Energy, Office of Science, Office of Basic Energy Sciences, under Contract No. DE-AC02-06CH11357.
\end{acknowledgments}

\textbf{Disclaimer}: The identification of any commercial product or a trade name does not necessarily imply endorsement or recommendation by the National Institute of Standards and Technology.


%

\end{document}